\documentclass[aps,prl,superscriptaddress,reprint]{revtex4-1}
\usepackage{amsmath}
\usepackage{amssymb}
\usepackage{graphicx}
\usepackage[colorlinks,urlcolor=blue]{hyperref}
\usepackage{url}
\usepackage{color,xcolor}
\usepackage{ulem}
\usepackage{float}
\usepackage{epstopdf}
\usepackage[none]{hyphenat}
\begin{document}

\title{Trends of electronic structures and $s_{\pm}$-wave pairing for the rare-earth series
\\in bilayer nickelate superconductor $R_ 3$Ni$_2$O$_7$}
\author{Yang Zhang}
\author{Ling-Fang Lin}
\affiliation{Department of Physics and Astronomy, University of Tennessee, Knoxville, Tennessee 37996, USA}
\author{Adriana Moreo}
\affiliation{Department of Physics and Astronomy, University of Tennessee, Knoxville, Tennessee 37996, USA}
\affiliation{Materials Science and Technology Division, Oak Ridge National Laboratory, Oak Ridge, Tennessee 37831, USA}
\author{Thomas A. Maier}
\affiliation{Computational Sciences and Engineering Division, Oak Ridge National Laboratory, Oak Ridge, Tennessee 37831, USA}
\author{Elbio Dagotto}
\affiliation{Department of Physics and Astronomy, University of Tennessee, Knoxville, Tennessee 37996, USA}
\affiliation{Materials Science and Technology Division, Oak Ridge National Laboratory, Oak Ridge, Tennessee 37831, USA}
\date{\today}

\begin{abstract}
The recent discovery of pressure-induced superconductivity in the bilayer La$_3$Ni$_2$O$_7$ (LNO) has opened a new platform for the study of unconventional superconductors. In this publication, we investigate theoretically the whole family of bilayer 327-type nickelates $R_3$Ni$_2$O$_7$ ($R$ = Rare-earth elements) under pressure. From La to Lu, the lattice constants and volume decrease, leading to enhanced in-plane and out-of-plane hoppings, resulting in an effectively reduced electronic correlation $U/W$. Furthermore, the Ni's $t_{2g}$ states shift away from the $e_g$ states, while the crystal-field splitting between $d_{3z^2-r^2}$ and $d_{x^2-y^2}$ is almost unchanged. In addition, six candidates were found to become stable in the Fmmm phase, with increasing values of critical pressure as the atomic number increases. Similar to the case of LNO, the $s_{\pm}$-wave pairing tendency dominates in all candidates, due to the nesting between the {\bf M}=$(\pi,\pi)$ and the {\bf X}=$(\pi,0)$ and {\bf Y}=$(0,\pi)$ points in the Brillouin zone. Then, $T_c$ is expected to decrease as the radius of rare-earth (RE) ions decreases. Our results suggest that LNO is already the ``optimal'' candidate, with Ce a close competitor, among the whole RE bilayer nickelates, and to increase $T_c$ we suggest to grow on special substrates with larger in-plane lattice spacings.
\end{abstract}

\maketitle
\section{I. Introduction}
The discovery of superconductivity in the bilayered perovskite La$_3$Ni$_2$O$_7$ (LNO) under high pressure~\cite{Sun:arxiv} provides a new platform for the study of nickelate-based superconductors. Different from the previously studied infinite-layer (IL) nickelate with the $d^9$ configuration~\cite{Li:Nature,Botana:prx,Nomura:rpp,Zhang:prb20,Gu:innovation}, LNO  has a $d^{\rm 7.5}$ configuration and a Ruddlesden-Popper (RP) perovskite structure involving NiO$_6$ bilayers, representing the first non-IL NiO$_2$ layered nickelate superconductor~\cite{Sun:arxiv}.

\begin{figure*}
\centering
\includegraphics[width=0.88\textwidth]{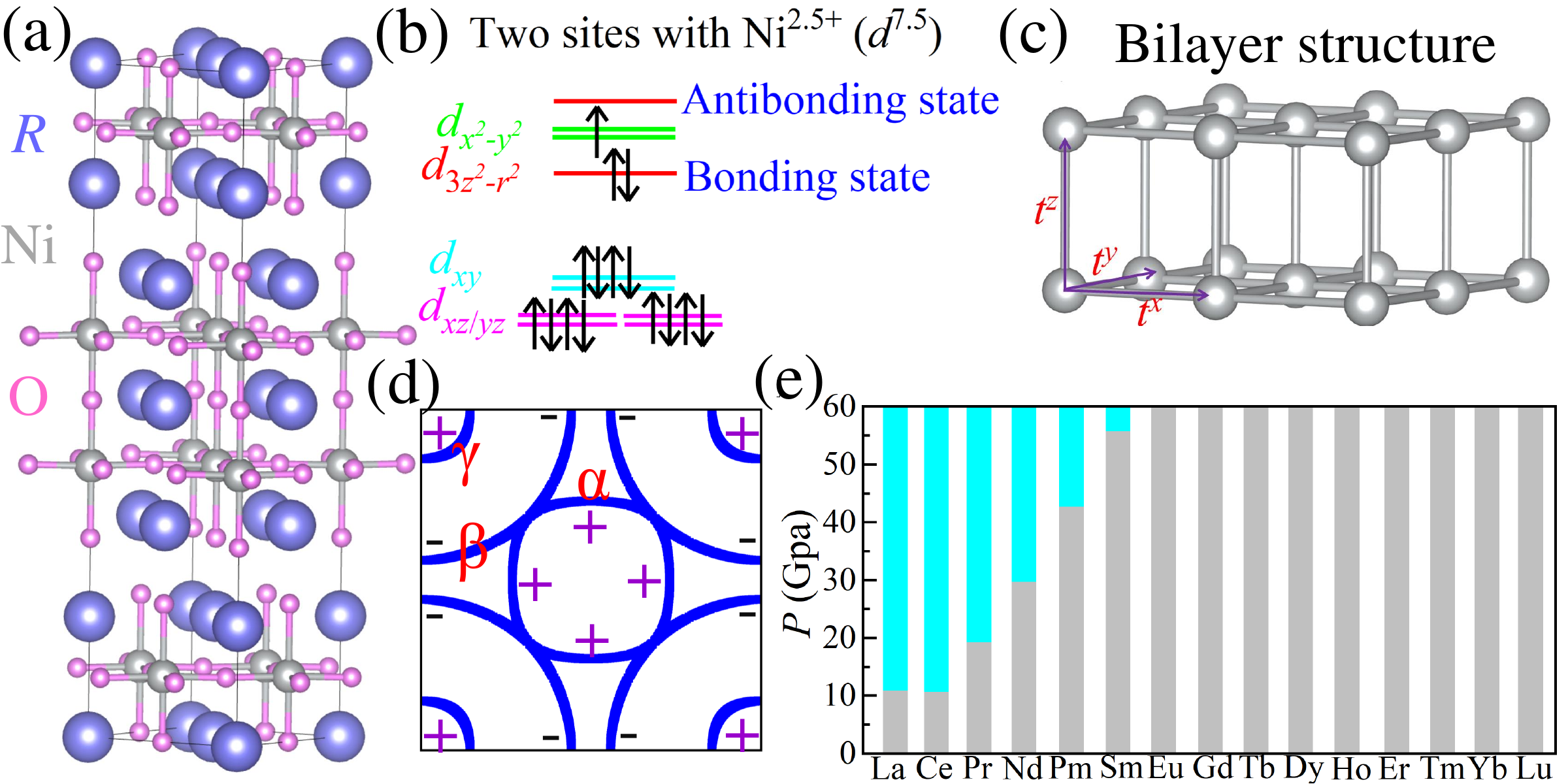}
\caption{(a) Schematic crystal structure of the conventional cell of $R_3$Ni$_2$O$_7$ ($R$ = RE elements) in the Fmmm phase (light blue = $R$; gray = Ni; pink = O). (b) Sketch of LNO electronic states with $d^{7.5+}$ in the small electronic correlation $U$ and $J$ region. Here, two Ni sites with 15 electrons are considered. The $d_{3z^2-r^2}$ orbitals form bonding-antibonding MO states while the $t_{2g}$ orbitals are fully occupied. (c) Sketch of the Ni bilayer structure. The local $z$-axis is perpendicular to the NiO$_6$ plane towards the top O atom, while the local $x$- or $y$-axis are along the in-plane Ni-O bond directions, leading to $d_{x^2-y^2}$ orbitals lying on the plane. (d) Sketch of LNO FS in the Fmmm phase, including a hole pocket at $\gamma$ and standard FS sheets at $\alpha$ and $\beta$, as well as the signs of the superconducting order parameter. (e) The stability of $R_3$Ni$_2$O$_7$ ($R$ = RE elements) at different pressures (0 to 60 Gpa). The cyan and gray represent whether the Fmmm phase is stable or not for different RE elements obtained from the phonon spectrum calculations. The boundary between cyan and gray indicates the critical pressure. All crystal structures in Fig.~\ref{F1}(a) and (c) were visualized with the VESTA code~\cite{Momma:vesta}.}
\label{F1}
\end{figure*}

Applying high pressure, LNO transforms from the Amam~\cite{Sun:arxiv,Liu:scpma} to the Fmmm structure (see Fig.~\ref{F1}(a)) where superconductivity was reported in a broad pressure range from 14 to 43.5 Gpa, with highest $T_c$ up to 80 K~\cite{Sun:arxiv}. Density functional theory (DFT) calculations revealed that the Fermi surface (FS) is contributed by the Ni's $d_{x^2-y^2}$ and $d_{3z^2-r^2}$ orbitals~\cite{Luo:arxiv,Zhang:arxiv}, while $t_{2g}$ orbitals are fully occupied. In addition, the Ni $d_{3z^2-r^2}$ orbital forms a bonding-antibonding molecular-orbital (MO) state~\cite{Sun:arxiv,Zhang:arxiv}, while the $d_{x^2-y^2}$ orbital is partially occupied (Fig.~\ref{F1}(b)), leading to an orbital-selective state~\cite{Zhang:arxiv}. Thus, this system can be described via a Ni two-orbital bilayer model~\cite{Luo:arxiv,Zhang:arxiv}, as in Fig.~\ref{F1}(c).

Moreover, the FS consists of two sheets with mixed $e_g$ orbitals ($\alpha$ and $\beta$)
and a hole-pocket $\gamma$ dominated by the $d_{3z^2-r^2}$ orbital~\cite{Luo:arxiv,Zhang:arxiv}, as displayed in Fig.~\ref{F1}(d). The partial nesting of the FS suggested that $s_{\pm}$-wave pairing superconductivity may be induced by spin fluctuations in the Fmmm phase of LNO~\cite{Yang:arxiv,Sakakibara:arxiv,Gu:arxiv,Shen:arxiv,Liu:arxiv,Zhang:arxiv1}. Furthermore, random-phase approximation (RPA) many-body calculations also suggest the key importance of the $\gamma$ hole pocket at wavevector $(\pi,\pi)$ in mediating superconductivity in LNO~\cite{Zhang:arxiv1}. This pocket vanishes for the non-superconducting
Amam phase of LNO~\cite{Sun:arxiv}. In addition, several groups also used the bilayer $t-J$ model to study the superconducting pairing symmetry~\cite{Lu:arxiv,Oh:arxiv,Liao:arxiv,Qu:arxiv,Yang:arxiv1}. Furthermore, the role of the Hund coupling~\cite{Zhang:arxiv,Cao:arxiv} and electronic correlation effects~\cite{Lechermann:arxiv,Christiansson:arxiv,LiuZhe:arxiv,Wu:arxiv,Shilenko:arxiv,Chen:arxiv} were also investigated in LNO. Very recently, a first-order transition under pressure~\cite{Zhang:arxiv1,Zhang:arxiv-exp,Hou:arxiv} and strange metal behavior were also discussed~\cite{Zhang:arxiv-exp}, with the structural phase transition suppressing superconductivity~\cite{Zhang:arxiv-exp,Hou:arxiv}.

Considering the various previously studied rare-earth (RE) elements in the IL nickelate and cuprates~\cite{Osada:nl,Hor:prl}, such as the $T_c$ being reduced by the replacement of Nd ($\sim 14$ K)~\cite{Li:Nature} by the larger ion-radius Pr ($\sim 7-12$ K)~\cite{Osada:nl} in IL nickelates and the $T_c$ being almost unchanged in the cuprates superconductor by the replacement of RE elements~\cite{Hor:prl}, then the next natural step is to replace La by other RE elements in LNO. Interesting questions naturally arise: could any other RE bilayer nickelates be stable in the Fmmm phase? What is the role of the RE element? Could superconductivity in the Fmmm phase develop, as in LNO? Will $T_c$ increase by RE replacement?

\section{II. Results}
\subsection{A. The effect of rare-earth element}
To answer these questions, we studied theoretically various RE elements replacing La, followed by the influence of
pressure, by using first-principles DFT~\cite{Kresse:Prb,Kresse:Prb96,Blochl:Prb,Perdew:Prl,Dudarev:prb} and the RPA calculations~\cite{Kubo2007,Graser2009,Altmeyer2016,Romer2020,Maier2022}. To understand the structural stability of the Fmmm phase of the bilayer RE nickelates $R_3$Ni$_2$O$_7$ ($R$ = RE elements), we calculated the phonon spectrum of the Fmmm phases under pressure, using the density functional perturbation theory approach~\cite{Baroni:Prl,Gonze:Pra1,Gonze:Pra2} analyzed by the PHONONPY software~\cite{Chaput:prb,Togo:sm}. Unfortunately, no RE bilayer nickelates was found to be stable in the Fmmm phase at 0 Gpa~\cite{Supplemental}. As displayed in Fig.~\ref{F1}(e), six RE candidates were found to become stable in the Fmmm phase in the pressure range we studied (0 to 60 Gpa): LNO with critical pressure $\sim 10.6$ Gpa~\cite{Zhang:arxiv1}, Ce$_3$Ni$_2$O$_7$ with critical pressure $\sim 11$ Gpa, Pr$_3$Ni$_2$O$_7$ with critical pressure $\sim 19$ Gpa, Nd$_3$Ni$_2$O$_7$ with critical pressure $\sim 30$ Gpa, Pm$_3$Ni$_2$O$_7$ with critical pressure $\sim 43$ Gpa, and Sm$_3$Ni$_2$O$_7$ with critical pressure $\sim 56$ Gpa. Next, we will mainly focus on the results of 30 Gpa to discuss the effect of different RE elements, unless otherwise specified.

\begin{figure}
\centering
\includegraphics[width=0.44\textwidth]{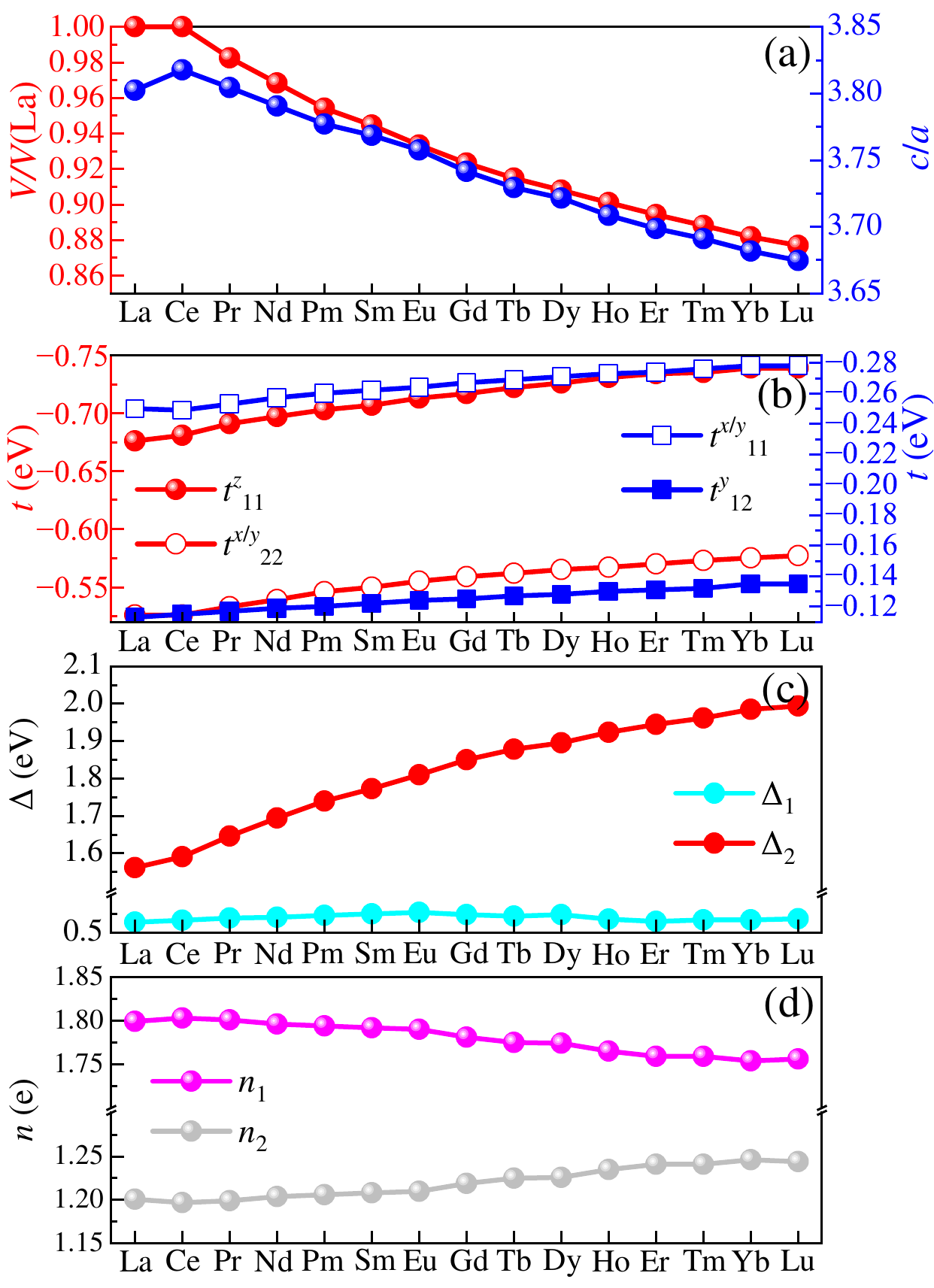}
\caption{(a) The change in volume compared with the optimized volume of LNO and the calculated ratio of $c/a$ at 30 Gpa for $R_3$Ni$_2$O$_7$ ($R$ = RE elements) at 30 Gpa. (b) The NN intraorbital and interorbital hoppings along $x$-, $y$-, and $z$-directions for $R_3$Ni$_2$O$_7$ ($R$ = RE elements) at 30 Gpa. The $\gamma = 1$ and $\gamma = 2$ orbitals correspond to $d_{3z^2-r^2}$ and $d_{x^2-y^2}$ orbitals, respectively. The intrahopping $t^{x}_{12}$ has the same value as $t^{y}_{12}$ but with an opposite sign (not shown here). (c) Crystal-field splittings $\Delta_1$ (between $d_{3z^2-r^2}$ and $d_{x^2-y^2}$ orbitals) and $\Delta_2$ (between $d_{3z^2-r^2}$ and $t_{2g}$ orbitals) for $R_3$Ni$_2$O$_7$ ($R$ = RE elements) at 30 Gpa. (d) The calculated electronic density for $R_3$Ni$_2$O$_7$ ($R$ = RE elements) at 30 Gpa. Here, the hopping and crystal field are obtained from the maximally localized Wannier functions~\cite{Mostofi:cpc} by fitting DFT and Wannier bands for different RE elements.
Note that these densities are for {\it two} sites, thus for the individual Ni atom occupancies results must
be divided by two.}
\label{RE}
\end{figure}

From La to Lu, the optimized out-of-plane and in-plane lattice constants are gradually reduced to about $6.4\%$ and $3.2\%$, respectively, of the ambient pressure values leading to the reduction of the volume by $\sim 12.3\%$, as well as the reduction of the ratio $c/a$ by about $3.2\%$~\cite{lattice}, as shown in Fig.~\ref{RE}(a). The changes in lattice constants and volume of the bilayer $R_3$Ni$_2$O$_7$ series are slightly smaller than the analogous changes in the IL nickelate $R$NiO$_2$ series ($\sim 10\%/5\%$ and $18.8\%$ for the $c$-axis/$a$-axis and volume, respectively)~\cite{Been:prx20}. Similar to LNO, the Ni $d_{3z^2-r^2}$ orbital forms a bonding-antibonding MO state and partially occupied $d_{3x^2-y^2}$ states, while the $t_{2g}$ orbitals are fully occupied for all RE elements, leading to an interesting orbital-selective phase~\cite{Zhang:arxiv,Zhang:ossp}.

As the radius of the RE ions decreases, the hopping $t^z_{11}$ ($t^z_{3z^2-r^2}$) increases about $9.3\%$, leading to a larger energy gap between the $d_{3z^2-r^2}$ bonding and antibonding states ($\Delta E$ $\sim$ 2$t^z_{11}$). Furthermore, the in-plane hopping of $d_{x^2-y^2}$ ($t^{x/y}_{22}$) and $d_{3z^2-r^2}$ ($t^{x/y}_{11}$) orbitals also increase by about $9.3\%$ and $9.7\%$, respectively, as shown in Fig.~\ref{RE}(b). These results are not surprising considering the in-plane and out-of-plane reduced lattice constants with increasing pressure that enhances the overlap of the orbitals between nearest-neighbor (NN) sites. In addition, the increased hoppings would also slightly increase the bandwidth $W$ of Ni's $e_g$ orbitals, leading to an ``effectively'' reduced electronic correlation $U/W$ from La to Lu bilayer nickelates.

In addition, the in-plane hybridization ($\lvert$$t_{12}/t_{22}$$\rvert$)between  $d_{3z^2-r^2}$ and $d_{x^2-y^2}$ is almost unchanged from La to Lu, as well as the crystal-field splitting $\Delta_1$ (between the $d_{3z^2-r^2}$ and $d_{x^2-y^2}$ orbitals), as shown in Fig.~\ref{RE}(c). However, the Ni's $t_{2g}$ states shift away from the $e_g$ states when RE changes from La to Lu, resulting in a large increase in $\Delta_2$ (between the $d_{3z^2-r^2}$ and $t_{2g}$ orbitals) of about $27.7\%$. Furthermore, the tight-binding (TB) calculations indicate that the ratio of electronic density of the $d_{3z^2-r^2}$ and $d_{x^2-y^2}$ are not changed much for different RE elements, as shown in Fig.~\ref{RE}(d). All these results suggest that the physical properties will not change drastically by replacing La by other RE elements. Hence, the $s_{\pm}$-wave pairing channel is to be expected for all RE cases.

\subsection{B. The pairing symmetry in the RE candidates}
Next, let us discuss the possible superconducting pairing symmetry in the $R_3$Ni$_2$O$_7$ series. First, we constructed a four-band orbital TB model in a bilayer lattice~\cite{Maier:prb11,Mishra:sr,Maier:prb19,Maier:prb22}, for the Fmmm phase of the various RE candidates~\cite{candidates}, involving two Ni sites with $e_g$ orbitals in a unit cell, by considering the overall filling $n = 3$. The kinetic hopping component is:
\begin{eqnarray}
H_k = \sum_{\substack{i\sigma\\\vec{\alpha}\gamma\gamma'}}t_{\gamma\gamma'}^{\vec{\alpha}}
(c^{\dagger}_{i\sigma\gamma}c^{\phantom\dagger}_{i+\vec{\alpha}\sigma\gamma'}+H.c.)+ \sum_{i\gamma\sigma} \Delta_{\gamma} n_{i\gamma\sigma}.
\end{eqnarray}
The first term represents the hopping of an electron from orbital $\gamma$ at site $i$ to orbital $\gamma'$ at the NN site $i+\vec{\alpha}$. $c^{\dagger}_{i\sigma\gamma}$($c^{\phantom\dagger}_{i\sigma\gamma}$) is the standard creation (annihilation) operator, $\gamma$ and $\gamma'$ represent the different orbitals, and $\sigma$ is the $z$-axis spin projection. $\Delta_{\gamma}$ represents the crystal-field splitting of each orbital $\gamma$. The unit vectors $\vec{\alpha}$ are along the three bilayer-lattice directions (see Fig.~\ref{F1}(c)).

\begin{figure}
\centering
\includegraphics[width=0.44\textwidth]{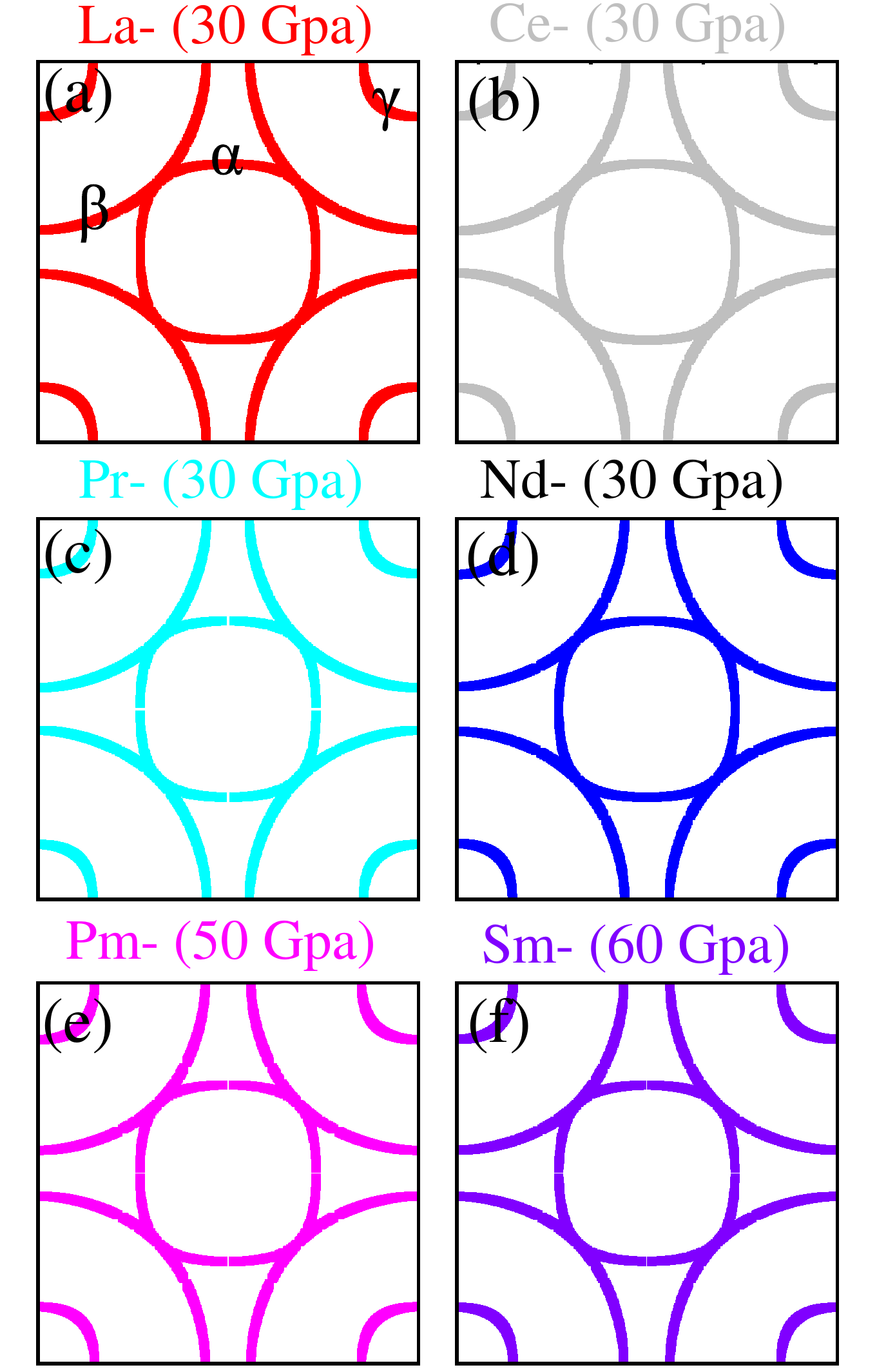}
\caption{(a-f) FSs of $R_3$Ni$_2$O$_7$ ($R$ = La to Sm) obtained from TB bilayer model calculations. (a) FS of LNO at 30 GPa. (b) FS of Ce$_3$Ni$_2$O$_7$ at 30 GPa. (c) FS of Pr$_3$Ni$_2$O$_7$ at 30 GPa. (d) FS of Nd$_3$Ni$_2$O$_7$ at 30 GPa. (e) FS of Pm$_3$Ni$_2$O$_7$ at 50 GPa. (f) FS of Sm$_3$Ni$_2$O$_7$ at 60 GPa.}
\label{FS}
\end{figure}

Because the states at the Fermi surface are related to the spin-fluctuation pairing interaction, we calculated the FSs for those six candidates. As shown in Fig.~\ref{FS}, the FSs of those RE candidates are nearly identical, consisting of two electron sheets ($\alpha$ and $\beta$) with a mixture of $d_{3z^2-r^2}$ and $d_{x^2-y^2}$ orbitals, while the $\gamma$ hole-pocket is made up of the $d_{3z^2-r^2}$ orbital. Note that here the Fermi surfaces seem almost identical, but they are slightly different from compound to compound. Specifically, while the hoppings, crystal fields, and electronic density are different from element to element, as shown
in the Appendix H, they are not sufficiently different to produce drastic changes in the Fermi surfaces. The size of the $\gamma$ pocket also slightly increases from La to Sm. For these reasons, the spin-fluctuation mediated $s^\pm$-wave pairing is also expected for the various RE candidates, as discussed before in the LNO case.

Next, we used the multi-orbital RPA model calculations to assess the bilayer TB models for those RE candidates to study superconducting behavior, based on a perturbative weak-coupling expansion in the Coulomb interaction~\cite{Romer2020,Kubo2007,Graser2009,Altmeyer2016,Maier2022}. The pairing strength $\lambda_\alpha$ for channel $\alpha$  and the corresponding gap structure $g_\alpha({\bf k})$ are obtained from solving an eigenvalue problem of the form:
\begin{eqnarray}\label{eq:pp}
	\int_{FS} d{\bf k'} \, \Gamma({\bf k -  k'}) g_\alpha({\bf k'}) = \lambda_\alpha g_\alpha({\bf k})\,,
\end{eqnarray}
where the momenta ${\bf k}$ and ${\bf k'}$ are on the FS, and $\Gamma({\bf k - k'})$ contains the irreducible particle-particle vertex. In the RPA approximation, the dominant term entering $\Gamma({\bf k-k'})$ is the RPA spin susceptibility $\chi({\bf k-k'})$.

 \begin{figure}
\centering
\includegraphics[width=0.44\textwidth]{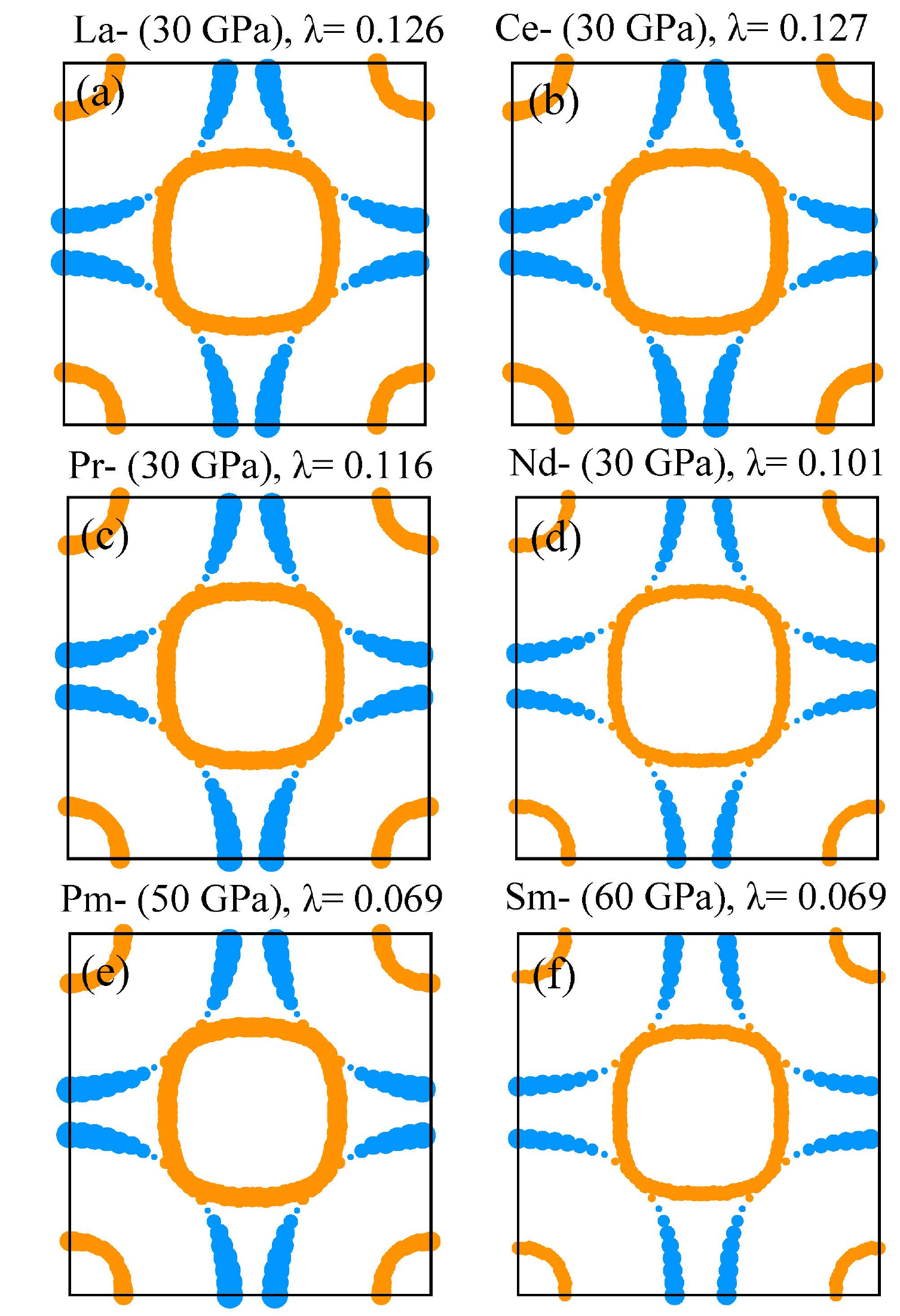}
\caption{RPA calculated leading superconducting gap structure $g_\alpha({\bf k})$ for momenta ${\bf k}$ on the FS for (a) La$_3$Ni$_2$O$_7$ at 30 GPa, (b) Ce$_3$Ni$_2$O$_7$ at 30 GPa, (c) Pr$_3$Ni$_2$O$_7$ at 30 GPa, (d) Nd$_3$Ni$_2$O$_7$ at 30 GPa, (e) Pm$_3$Ni$_2$O$_7$ at 50 GPa, and (f) Sm$_3$Ni$_2$O$_7$ at 60 GPa. They all display $s^\pm$ symmetry, where the gap changes sign between the $\alpha$ and $\beta$ FS sheets, and also between the $\beta$ sheet and $\gamma$ pocket. The sign of the gap is indicated by color (orange = positive, blue = negative), and gap amplitude by thickness. Here we used $U=0.8$, $U'=0.4$, and $J=J'=0.2$ in units of eV, as in our previous work~\cite{Zhang:arxiv1}. All hoppings and crystal fields for different RE elements are in the Fmmm phase.}
\label{GAP}
\end{figure}

Figure~\ref{GAP} displays the leading pairing symmetry $g_{\alpha}({\bf k})$ obtained from solving the eigenvalue problem in Eq.~(\ref{eq:pp}) for the RPA pairing interaction of $R_3$Ni$_2$O$_7$ ($R$ = La to Sm).
In these cases, we found the $s^\pm$ gap structure is always the leading channel, caused by spin-fluctuation. The gap switches sign between the $\alpha$ and $\beta$ FS sheets, and between the $\beta$ sheet and $\gamma$ pocket (see Fig.~\ref{GAP}). Similar to our previous analysis for LNO under pressure~\cite{Zhang:arxiv1}, the gap structures of $R_3$Ni$_2$O$_7$ ($R$ = La to Sm) are driven by intra-orbital $(\pi, 0)$ scattering between the $(0,\pi)$ region of the $\beta$-sheet with significant $d_{3z^2-r^2}$ character of the Bloch states, to the $d_{3z^2-r^2}$ $\gamma$-pocket at $(\pi,\pi)$. In addition, the $s^\pm$ gap remains the leading instability over the subleading $d_{x^2-y^2}$ gap for the six cases we considered.

Furthermore, within the RPA approximation, the superconducting pairing arises from spin fluctuations. Here, our RPA results indicate a magnetic stripe order,  where these fluctuations have a wave-vector $(\pi, 0)$ and $(0,\pi)$ and, therefore, scatter primarily between the $\beta$ electron Fermi surface sheet and the $\gamma$ hole pocket, which are partially connected by these vectors. In addition, very recent theoretical studies also suggest a possible spin stripe ($\pi$, 0) instability order in LNO~\cite{Luo:arxiv,Shilenko:arxiv,Zhang:arxiv,Zhang:arxiv1}, with a mixture of ferromagnetic (FM) bonds in one direction and antiferromagnetic (AFM) bonds in the other direction, which can be understood by the strong competition between intraorbital and interorbital hopping mechanisms~\cite{Lin:prl21,Lin:prb,Lin:cp}, namely to the competition of AFM and FM tendencies~\cite{Zhang:arxiv,Zhang:arxiv1}. Although no long-range magnetic order was found experimentally in LNO yet, short-range magnetic spin order involving localized moments is still possible, similarly to the case experimentally observed in the IL nickelate as the quality of the samples improved~\cite{Li:cm,Fowlie:np}.

From La to Nd bilayer nickelates, the pairing strength $\lambda$ slightly decreases from 0.126 to 0.101. However, for the Pm- and Sm-bilayer nickelates, the pairing strength $\lambda$ is reduced more, by about $45\%$, because we considered much higher pressure values for these cases. The $(\pi,0)$ scattering is known to decrease with increasing pressure, as discussed in LNO~\cite{Zhang:arxiv1}. Furthermore, the pairing strength $\lambda$ can be linked to the superconducting transition temperature $T_c$ through a Bardeen-Cooper-Schrieffer (BCS) like equation, $T_c=\omega_0e^{-1/\lambda}$, where $\omega_0$ is a cut-off frequency determined from the spin-fluctuation spectrum.
Note here that the Matsubara energy dependence was not considered in our RPA calculation, even though an inclusion of the Matsubara frequency dependence is numerically more accurate. However, it is also more numerically demanding, because all the states must be taken into account. Our approach is based on the BCS assumption that the spin-fluctuation pairing interaction is approximately independent of energy in a shell around the Fermi surface up to an energy cut-off $\omega_0$ and given by its static limit. Using this assumption, the Matsubara sum in the Eliashberg equations can be performed analytically, and one arrives at the equations we provide in the paper in which only the states on the Fermi surface are considered. This procedure is discussed in more detail in Ref.~\cite{Bhattacharyya:prb}. It is much less demanding than solving the full Eliashberg equations and has been routinely used in the literature, especially in the context of iron-based superconductivity. In the present problem, there are no other low-lying states or incipient bands that are not represented by the Fermi surface states present in the electronic structure. Therefore, we do not expect a treatment that includes energy dependence, and such potential states, to give qualitatively different results.

In addition, within RPA in general we find that the $T_c$ increases as the pressure decreases in LNO~\cite{Zhang:arxiv1}. Considering that the same $s_{\pm}$-wave pairing and similar electronic structures were obtained in our present work by changing the chemical formula varying RE elements, it is then reasonable that $T_c$ is expected to decrease as the radius of the RE ions decreases. As discussed in Fig.~\ref{F1}(e), the tendency of the Amam-Fmmm phase boundary clearly shows that it needs much higher pressure to stabilize the Fmmm phase as the RE number increases. Then, it appears that LNO is already close to being the ``optimal'' candidate, with Ce a close competitor, in the bilayer 327-type family from our RPA perspective.

\section{III. Additional Discussions}
The phase diagram of this LNO system under pressure is quite different from the previously known phase diagram of Cu-based and Fe-based superconductors under pressure: we do not find an obvious sharp and narrow superconducting dome, but instead superconductivity appears in a very broad pressure region~\cite{Sun:arxiv,Zhang:arxiv-exp}. Furthermore, the structural phase transition from the Amam to Fmmm phases was also reported at some regions of lower pressures~\cite{Sun:arxiv,Zhang:arxiv-exp,Hou:arxiv} due to its first order nature~\cite{Zhang:arxiv-exp,Hou:arxiv}. Very recent theoretical studies indicate that the superconducting pairing strength $\lambda$ is dramatically reduced to zero without the $\gamma$ pocket~\cite{Zhang:arxiv1}, while this pocket vanishes in the non-superconducting Amam phase~\cite{Sun:arxiv}. In this case, it is reasonable to assume a similar superconducting phase diagram under pressure for other RE elements. Hence, to study the pairing strength $\lambda$ varying pressure for different rare-earth elements, especially for a very dense grid of pressures, is a very interesting future work that deserves further investigation.

Compared with other RE elements, our work also suggests that LNO seems to be an ``optimal'' candidate in the bilayer 327-type family from our theoretical perspective, with Ce a close competitor. Typically, tetragonal substrates would suppress octahedral distortions. Consequently, to obtain a higher $T_c$, in-plane tensile stress-strain or growing on substrates with {\it larger in-plane lattice constants} would be the direction to stabilize the Fmmm phase at lower or ambient pressures. In other words, by artificially engineering larger $a$ and $b$ lattice spacings, the material may move {\it opposite} to the direction spontaneously selected by Rare Earth replacement, as described in the present publication, hopefully reducing the critical pressure needed to reach the Fmmm phase. Note, however, that via substrate growth, increasing $a$ and $b$ will likely result in a smaller $c$, and thus this matter needs careful further studies that we will present in near future work.

In addition, we also calculated the phonon spectrum of the Fmmm phases under pressure for two lighter RE elements Sc and Y~\cite{Supplemental}. However, none of them was found to be stable in the Fmmm phase, at least in the pressure region we studied.

\section{IV. Conclusions}
In summary, here we comprehensively studied the effect of RE elements in the Fmmm phase of the bilayer 327-type nickelates under pressure. In the explored pressure region (0 to 60 Gpa), the Fmmm phase was found to become stable in six RE candidates. In those six candidates, from La to Sm, the critical value of pressure for stabilizing the Fmmm phase increases. To stabilize the Fmmm phase in the other RE element bilayer nickelate systems, a much higher pressure is needed. Furthermore, as discussed in our previous work~\cite{Zhang:arxiv1}, the Amam and Fmmm phases could be coexisting in a small range of pressure, due to the first-order structural transition nature. Hence, the critical value of pressure for pure Fmmm phases would be effectively even higher. From La to Lu, the optimized out-of-plane and in-plane lattice constants are gradually reduced, leading to slightly increased $e_g$ orbital hoppings. Furthermore, the crystal-field splitting $\Delta_1$ between $d_{3z^2-r^2}$ and $d_{x^2-y^2}$ orbitals does not change much as the radius of the RE ions decreases. However, the $\Delta_2$ between the $d_{3z^2-r^2}$ and $t_{2g}$ orbitals increases by larger values.

Furthermore, we also calculated the FSs of all the explored RE candidates, using a four-band $e_g$ bilayer model. Similarly to the case of LNO, we found two electron FS sheets ($\alpha$ and $\beta$) with a mixture of $d_{3z^2-r^2}$ and $d_{x^2-y^2}$ orbitals, and a $\gamma$ hole-pocket made up of the $d_{3z^2-r^2}$ orbital. For this reason, our RPA results show the same spin-fluctuation mediated $s^\pm$-wave pairing in the RE candidates studied. However, the pairing strength $\lambda$ is reduced from La to Sm, suggesting $T_c$ is reduced as the radius of the RE ions decreases.

\section{V. Acknowledgments}
This work was supported by the U.S. Department of Energy, Office of Science, Basic Energy Sciences, Materials Sciences and Engineering Division.

\section{VI. APPENDIX}

\subsection{A. Method}
In this work, first-principles DFT calculations were implemented via the Vienna {\it ab initio} simulation package (VASP) code, by using the projector augmented wave (PAW) method~\cite{Kresse:Prb,Kresse:Prb96,Blochl:Prb}, within
the generalized gradient approximation and the Perdew-Burke-Ernzerhof exchange potential~\cite{Perdew:Prl}. In addition, we treat the $4f$ electrons in the core, as discussed in the infinite-layer $R$NiO$_2$ ($R$ = RE elements)~\cite{Been:prx20}.
Furthermore, the plane-wave cutoff energy was set as $550$~eV and a $k$-point grid $12\times12\times3$ was used for the conventional structure of $R_3$Ni$_2$O$_7$ ($R$ = RE elements) of the Fmmm phases under pressure.

Here, we considered a pressure grid with an
interval of 5 Gpa for different RE elements, but the interval changed to 1 Gpa near the critical pressures. Furthermore, both lattice constants and atomic positions were fully relaxed until the Hellman-Feynman force on each atom was smaller than $0.01$ eV/{\AA}. Moreover, the onsite Coulomb interactions were considered via the Dudarev formulation~\cite{Dudarev:prb} with $U_{eff} = 4$ eV, as used in recent studies of La$_3$Ni$_2$O$_7$~\cite{Sun:arxiv,Christiansson:arxiv,Chen:arxiv,Zhang:arxiv1}.

To investigate the structural stability of the Fmmm phase, we studied the phonon spectra for different pressures of the bilayer RE nickelates $R_3$Ni$_2$O$_7$ in the conventional cell structures, by using the density functional perturbation theory approach~\cite{Baroni:Prl,Gonze:Pra1,Gonze:Pra2}, analyzed by the PHONONPY software in the primitive unit cell~\cite{Chaput:prb,Togo:sm}.
To avoid repeating too many displays, we only show the phonon spectrum of several key values of pressures for different RE elements in the Supplemental Materials~\cite{Supplemental}.

In addition to the standard DFT calculation discussed thus far, the maximally localized Wannier functions (MLWFs) method was employed to fit Ni's $e_g$ bands to obtain the hoppings and crystal-field splittings for our subsequent model calculations, using the WANNIER90 packages~\cite{Mostofi:cpc}.

Furthermore, we also considered the many-body RPA to study the bilayer TB models for their superconducting behavior, based on a perturbative weak-coupling expansion in the Coulomb interaction~\cite{Kubo2007,Graser2009,Altmeyer2016,Romer2020,Maier2022} approach, which has been already shown in many studies to capture the essence of the physics (e.g. Ref.~\cite{Romer2020}). In the multi-orbital RPA technique~\cite{Kubo2007,Graser2009,Altmeyer2016}, the RPA enhanced spin susceptibility is obtained from the bare susceptibility (Lindhart function) $\chi_0({\bf q})$ as $\chi({\bf q}) = \chi_0({\bf q})[1-{\cal U}\chi_0({\bf q})]^{-1}$. Here, $\chi_0({\bf q})$ is an orbital-dependent susceptibility tensor and ${\cal U}$ is a tensor that contains the intra-orbital $U$ and inter-orbital $U'$ density-density interactions, the Hund's rule coupling $J$, and the pair-hopping $J'$ term. The pairing strength $\lambda_\alpha$ for channel $\alpha$ and the corresponding gap structure $g_\alpha({\bf k})$ are obtained from solving an eigenvalue problem of the form in Eq.~\ref{eq:pp}. More details of the multi-orbital RPA approach can be found in \cite{Kubo2007,Graser2009,Altmeyer2016,Maier2022}.

\subsection{B. Hoppings for La$_3$Ni$_2$O$_7$ at 30 Gpa.}

The hopping matrix (in eV units) for La$_3$Ni$_2$O$_7$ at 30 Gpa. $\Delta_1 = 0.528$ eV.

\begin{equation}
\begin{split}
t_{\vec{x}} =
\begin{bmatrix}
          d_{z^2}   &      d_{x^2-y^2}   \\
         -0.113  	&        0.250	   \\
          0.250     &       -0.526	  	
\end{bmatrix},\\
\end{split}
\label{hopping1}
\end{equation}

\begin{equation}
\begin{split}
t_{\vec{y}} =
\begin{bmatrix}
          d_{z^2}   &      d_{x^2-y^2}   \\
         -0.113  	&       -0.250	   \\
         -0.250	    &       -0.526	  	
\end{bmatrix},\\
\end{split}
\label{hopping1}
\end{equation}

\begin{equation}
\begin{split}
t_{\vec{z}} =
\begin{bmatrix}
          d_{z^2}   &      d_{x^2-y^2}   \\
         -0.676  	&        0.000	   \\
          0.000	    &        0.000	  	
\end{bmatrix},\\
\end{split}
\label{hopping1}
\end{equation}

\subsection{C. Hoppings for Ce$_3$Ni$_2$O$_7$ at 30 Gpa.}
The hopping matrix (in eV units) for Ce$_3$Ni$_2$O$_7$ at 30 Gpa. $\Delta_1 = 0.533$ eV.

\begin{equation}
\begin{split}
t_{\vec{x}} =
\begin{bmatrix}
          d_{z^2}   &      d_{x^2-y^2}   \\
         -0.115 	&        0.249	   \\
          0.249     &       -0.526	  	
\end{bmatrix},\\
\end{split}
\label{hopping1}
\end{equation}

\begin{equation}
\begin{split}
t_{\vec{y}} =
\begin{bmatrix}
          d_{z^2}   &      d_{x^2-y^2}   \\
         -0.115  	&       -0.249	   \\
         -0.249	    &       -0.526	  	
\end{bmatrix},\\
\end{split}
\label{hopping1}
\end{equation}

\begin{equation}
\begin{split}
t_{\vec{z}} =
\begin{bmatrix}
          d_{z^2}   &      d_{x^2-y^2}   \\
         -0.681  	&        0.000	   \\
          0.000	    &        0.000	  	
\end{bmatrix},\\
\end{split}
\label{hopping1}
\end{equation}

\subsection{D. Hoppings for Pr$_3$Ni$_2$O$_7$ at 30 Gpa.}
The hopping matrix (in eV units) for Pr$_3$Ni$_2$O$_7$ at 30 Gpa. $\Delta_1 = 0.539$ eV.

\begin{equation}
\begin{split}
t_{\vec{x}} =
\begin{bmatrix}
          d_{z^2}   &      d_{x^2-y^2}   \\
         -0.117 	&        0.253	   \\
          0.253     &       -0.533	  	
\end{bmatrix},\\
\end{split}
\label{hopping1}
\end{equation}

\begin{equation}
\begin{split}
t_{\vec{y}} =
\begin{bmatrix}
          d_{z^2}   &      d_{x^2-y^2}   \\
         -0.117  	&       -0.253	   \\
         -0.253	    &       -0.533	  	
\end{bmatrix},\\
\end{split}
\label{hopping1}
\end{equation}

\begin{equation}
\begin{split}
t_{\vec{z}} =
\begin{bmatrix}
          d_{z^2}   &      d_{x^2-y^2}   \\
         -0.691  	&        0.000	   \\
          0.000	    &        0.000	  	
\end{bmatrix},\\
\end{split}
\label{hopping1}
\end{equation}

\subsection{E. Hoppings for Nd$_3$Ni$_2$O$_7$ at 30 Gpa.}
The hopping matrix (in eV units) for Nd$_3$Ni$_2$O$_7$ at 30 Gpa. $\Delta_1 = 0.541$ eV.

\begin{equation}
\begin{split}
t_{\vec{x}} =
\begin{bmatrix}
          d_{z^2}   &      d_{x^2-y^2}   \\
         -0.119 	&        0.257	   \\
          0.257     &       -0.539	  	
\end{bmatrix},\\
\end{split}
\label{hopping1}
\end{equation}

\begin{equation}
\begin{split}
t_{\vec{y}} =
\begin{bmatrix}
          d_{z^2}   &      d_{x^2-y^2}   \\
         -0.119  	&       -0.257	   \\
         -0.257	    &       -0.539	  	
\end{bmatrix},\\
\end{split}
\label{hopping1}
\end{equation}

\begin{equation}
\begin{split}
t_{\vec{z}} =
\begin{bmatrix}
          d_{z^2}   &      d_{x^2-y^2}   \\
         -0.697  	&        0.000	   \\
          0.000	    &        0.000	  	
\end{bmatrix},\\
\end{split}
\label{hopping1}
\end{equation}

\subsection{F. Hoppings for Pr$_3$Ni$_2$O$_7$ at 50 Gpa.}
The hopping matrix (in eV units) for Pr$_3$Ni$_2$O$_7$ at 50 Gpa. $\Delta_1 = 0.587$ eV.

\begin{equation}
\begin{split}
t_{\vec{x}} =
\begin{bmatrix}
          d_{z^2}   &      d_{x^2-y^2}   \\
         -0.131 	&        0.277	   \\
          0.277     &       -0.580	  	
\end{bmatrix},\\
\end{split}
\label{hopping1}
\end{equation}

\begin{equation}
\begin{split}
t_{\vec{y}} =
\begin{bmatrix}
          d_{z^2}   &      d_{x^2-y^2}   \\
         -0.131  	&       -0.271	   \\
         -0.277	    &       -0.580	  	
\end{bmatrix},\\
\end{split}
\label{hopping1}
\end{equation}

\begin{equation}
\begin{split}
t_{\vec{z}} =
\begin{bmatrix}
          d_{z^2}   &      d_{x^2-y^2}   \\
         -0.750  	&        0.000	   \\
          0.000	    &        0.000	  	
\end{bmatrix},\\
\end{split}
\label{hopping1}
\end{equation}

{\subsection{G. Hoppings for Sm$_3$Ni$_2$O$_7$ at 60 Gpa.}
The hopping matrix (in eV units) for Pr$_3$Ni$_2$O$_7$ at 50 Gpa. $\Delta_1 = 0.618$ eV.

\begin{equation}
\begin{split}
t_{\vec{x}} =
\begin{bmatrix}
          d_{z^2}   &      d_{x^2-y^2}   \\
         -0.137 	&        0.288	   \\
          0.288     &       -0.598	  	
\end{bmatrix},\\
\end{split}
\label{hopping1}
\end{equation}

\begin{equation}
\begin{split}
t_{\vec{y}} =
\begin{bmatrix}
          d_{z^2}   &      d_{x^2-y^2}   \\
         -0.137  	&       -0.288	   \\
         -0.288	    &       -0.598	  	
\end{bmatrix},\\
\end{split}
\label{hopping1}
\end{equation}

\begin{equation}
\begin{split}
t_{\vec{z}} =
\begin{bmatrix}
          d_{z^2}   &      d_{x^2-y^2}   \\
         -0.779  	&        0.000	   \\
          0.000	    &        0.000	  	
\end{bmatrix},\\
\end{split}
\label{hopping1}
\end{equation}

\subsection{H. TB band structures for six candidates}
Furthermore, we also showed the TB band structures of those six candidates in the main text, to better display the (small) differences for different rare-earth elements.

\begin{figure}
\centering
\includegraphics[width=0.44\textwidth]{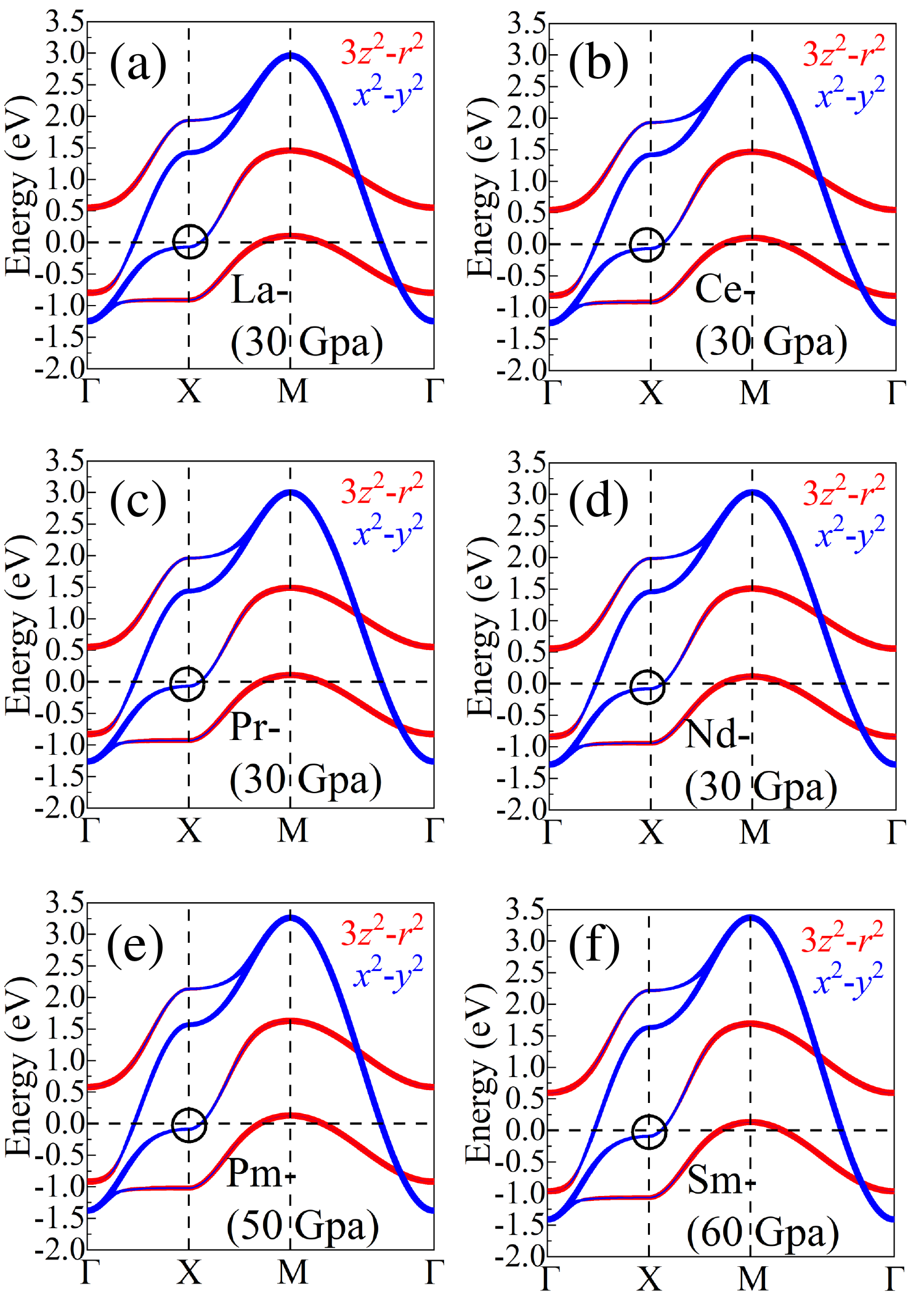}
\caption{(a-f) Band structures of $R_3$Ni$_2$O$_7$ ($R$ = La to Sm) obtained from TB bilayer model calculations. (a) FS of LNO at 30 GPa. (b) FS of Ce$_3$Ni$_2$O$_7$ at 30 GPa. (c) FS of Pr$_3$Ni$_2$O$_7$ at 30 GPa. (d) FS of Nd$_3$Ni$_2$O$_7$ at 30 GPa. (e) FS of Pm$_3$Ni$_2$O$_7$ at 50 GPa. (f) FS of Sm$_3$Ni$_2$O$_7$ at 60 GPa. Furthermore, the van Hove singularity near the X point is also marked by black circles.}
\label{Pband-TB}
\end{figure}


\begin{references}
\bibitem{lattice} {Compared to the La-case, in the Ce-case the optimized $c$-axis lattice increases by about $0.3\%$, while the calculated in-plane lattices decrease by about $0.1\%$, leading to an increased ratio $c/a$. Furthermore, the tendency of the changes in lattice constants and volume of the bilayer $R_3$Ni$_2$O$_7$ series are independent of the values of pressure. For example, at 0 Gpa, the calculated $a$ and $c$ lattice constants gradually reduces to about $6.4\%$ or $2.9\%$, respectively, leading to the reduction of the volume by $\sim 11.7\%$, as well as the reduction of the ratio $c/a$ by about $3.6\%$.}
\bibitem{candidates} {The Fmmm phase of Pm$_3$Ni$_2$O$_7$ and Sm$_3$Ni$_2$O$_7$ are considered to be stable above 43 and 56 Gpa, respectively. Thus, we used the hopping matrix and crystal field splitting of Pm$_3$Ni$_2$O$_7$ and Sm$_3$Ni$_2$O$_7$ at 50 and 60 Gpa, respectively. All the hopping matrix and crystal field splitting $\Delta_2$ can be found in the Appendix.}
\end{references}
\end{document}